\documentclass[amsmath,amssymb,twocolumn,prb
]{revtex4-2}

\usepackage{caption}

\usepackage{verbatim}
\usepackage{subcaption}
\usepackage{graphicx}
\usepackage{dcolumn}
\usepackage{bm}
\usepackage{mathrsfs}
\usepackage{hyperref}

\usepackage{xcolor}

\newcommand{\pvec}[1]{\vec{#1}\mkern2mu\vphantom{#1}}

\begin{document}
\title{Bulk Hydrodynamic Transport in Weyl Semimetals}

\author{Joan Bernabeu}
\email{joan.bernabeu@uam.es}
\affiliation{Departamento de Física de la Materia Condensada, Universidad Autónoma de Madrid, Cantoblanco, E-28049 Madrid, Spain}

\author{Kitinan Pongsangangan}
\email{kitinan.pon@mahidol.ac.th}
\affiliation{Department of Physics, Faculty of Science, Mahidol University, Bangkok 10400, Thailand}
\affiliation{Institute of Theoretical Physics, Technische Universität Dresden, 01062 Dresden, Germany}
%
\author{Henk T.C. Stoof}
\email{h.t.c.stoof@uu.nl}
\affiliation{Institute for Theoretical Physics and Center for Extreme Matter and Emergent Phenomena, Utrecht University, Princetonplein 5, 3584 CC Utrecht, The Netherlands}
\author{Lars Fritz}
\email{l.fritz@uu.nl}
\affiliation{Institute for Theoretical Physics and Center for Extreme Matter and Emergent Phenomena, Utrecht University, Princetonplein 5, 3584 CC Utrecht, The Netherlands}

\begin{abstract}
The role of collective longitudinal modes, plasmons, in bulk hydrodynamic transport in Weyl semimetals is explored. In contrast to graphene, where these modes are gapless, plasmons in Weyl semimetals are gapped. This gap, however, can be made arbitrarily small by decreasing the temperature or the chemical potential, making plasmon modes thermally accessible, both in thermodynamics and transport. In very clean Weyl semimetals near charge-neutrality where the plasmon gap is minimal, we find that they leave an imprint in the thermal conductivity and the viscosity. 
\end{abstract}

\maketitle
\section{Introduction}

Weyl semimetals (WSM) have emerged as a class of materials characterized by their unique electronic structure. Close to charge neutrality, these systems host Weyl fermions as low-energy quasiparticles \cite{wan2011topological,burkov2011weyl}. They arise from band crossings near the Fermi level that act as source terms of Berry curvature akin to magnetic monopoles. The resulting topology of these materials gives rise to a host of exotic phenomena, including unconventional surface states dubbed Fermi arcs \cite{wan2011topological,burkov2011weyl} and transport properties derived from the Adler-Bell-Jackiw chiral anomaly \cite{adler1969axial,bell1969PCAC}. The most well-known examples of the latter are the intrinsic anomalous Hall effect \cite{yang2011quantum,grushin2012consequences,zyuzin2012topological,vazifeh2013electromagnetic}, the chiral magnetic effect \cite{vazifeh2013electromagnetic,basar2014triangle} and its related negative longitudinal magnetoresistance \cite{2013_Son_Chiral,burkov2015chiral}. Furthermore, bulk thermoelectric transport in WSM provides a powerful lens into their fundamental properties \cite{hosur2012charge,lundgren2014thermoelectric,sharma2016nernst} as unlike conventional semimetals or metals, WSM can exhibit transport phenomena deeply influenced by their topological nature. Moreover, the coupling between bulk and surface states \cite{gorbar2016origin} can further convolute transport dynamics.

In parallel, the advent of ultra-clean two-dimensional (2D) materials such as graphene and other 2D systems has revolutionized the study of electronic transport in (weakly) interacting systems. The reduction of scattering due to impurities and the intrinsic suppression of phonon scattering has enabled the observation of hydrodynamic electronic transport—a long sought after transport regime in electronic systems. In that regime, the collective flow of electrons resembles the dynamics of classical fluids. Hallmarks of this behavior such as the Poiseuille flow or the minimal interaction-dominated conductivity have been observed experimentally. These developments have provided a fresh perspective on non-local transport phenomena and the interplay between ballistic, diffusive, and hydrodynamic regimes~\cite{borisreview1,borisreview2,lucas1,levchenko,lucas2,zaanen,geimpolini,kashuba,fritz,foster,mueller,schuett,narozhny,briskot,gallagher,tan,jonathan,vignale,wagner1,wagner2,tan,subir,levchenkoandreev,jonah,molenkamp,dejong,moll,nandi,sulpizio,gooth,ku}.

A first step to address the hydrodynamic regime in WSM, the three-dimensional analogues of graphene, was undertaken by Ref.~\cite{hosur2012charge}, where the authors determined the minimal interaction-dominated conductivity at charge neutrality in perturbation theory. A question that goes beyond this quantity when studying the hydrodynamic behavior of interacting electrons concerns the role of collective intrinsic and extrinsic quantum modes and their contribution to transport properties. For instance, electron-phonon interactions have been speculated to bring about a joint electron-phonon hydrodynamic regime \cite{Levchenko2020Transport,Huang2021Electron}, or a significant source of phonon-mediated electron-electron scattering needed for conventional electron hydrodynamics \cite{gurzhi1963minimum,Gurzhi1968hydro} in WSM such as WP$_2$ and WTe$_2$ \cite{Coulter2018Microscopic,Garcia2021Anisotropic,Vool2021Imaging,Bernabeu2023Bounds}.

Recently, some of the authors have suggested graphene to be a platform that could potentially realize the interplay of electrons and intrinsic quantum modes, plasmons \cite{Kitinan2022a,Kitinan2022b,Kitinan2022c,Kitinan2024}, in its hydrodynamic regime. In three dimensional Fermi liquids, plasmons have a large energy gap which leads to their absence in low-energy transport properties. In 2D (semi-)metallic systems, however, plasmons and electrons can coexist at the same energy scale. Some of the authors found that in the appropriate temperature range, plasmons provide a noticeable contribution to transport coefficients such as heat conductivity and shear viscosity \cite{Kitinan2022c,Kitinan2024}. 
The insights gained from 2D materials motivate similar explorations in WSM, where the interplay between topology, electron correlations, and disorder creates a rich landscape for unconventional transport phenomena. Indeed, WSM offer a three-dimensional playground where novel hydrodynamic and topological effects may coexist or compete \cite{lucas2016hydrodynamic,gorbar2018consistent,gorbar2018hydrodynamic,sukhachov2018collective}.

In this paper we concentrate on the conventional, non-anomalous, part of the relativistic hydrodynamic description, although we go beyond the standard case by taking collective modes into account. We set up a full `phenomenological' hydrodynamic description of interacting Weyl systems in the framework of the Boltzmann equation. From the experimental side, WSM with unprecedented high mobility, namely TaAs and NbAs, have already been realized and can be candidates for hydrodynamic behavior\cite{zhang2016taas,li2019nbas}. Furthermore, by means of high-resolution electron-energy loss spectroscopy, they were shown to exhibit absorption peaks corresponding to a plasmon-excitation energy gap of around $70$ meV at room temperature \cite{Chiarello2019}. The gap is tunable with temperature and doping and expected to reduce to 20 meV at charge neutrality \cite{2015_Hofman_Plasmon}. This contrasts with the aforementioned huge plasmon energy gap of $15-20$ eV for 3D metals \cite{Ashcroft&Mermin1976}. Thus, in WSM, we expect both plasmons and Dirac electrons to be excited at experimentally accessible temperatures. Analogous to graphene,  plasmons in WSM can be expected to show up in heat transport and shear viscosity.

The organization of this paper is as follows. We briefly explain  in Sec. \ref{sec:model} how to obtain, on the basis of the random-phase approximation, the electron-plasmon theory from electrons with long-range Coulomb interactions only. Furthermore, we sketch the derivation of coupled Boltzmann equations for this system within the non-equilibrium quantum field-theoretical method. The readers interested in comprehensive detail are recommended to Ref. \cite{Kitinan2022b}. In addition, we present an alternative effective-field-theory approach to derive these collective electromagnetic modes. Next, we solve in Sec.\ref{sec:transport} the coupled Boltzmann equations to study thermo-electric effects and the shear viscosity in the relaxation time approximation. Using certain relations between the relaxation times we manage to access the hydrodynamic regime and find in general an enhancement of the effects due to collective contributions. We finish in Sec.~\ref{discussion_section} with a conclusion and outlook. 

\section{Model}
\label{sec:model}
We consider a low-energy effective action for interacting Weyl fermions in the vicinity of a Weyl point. In a real material, Weyl points come in $N$ pairs of opposite chiralities \cite{NIELSEN198120,NIELSEN1981173}. As long as the interactions between electrons from different valleys are negligibly small, one can consider simply a single Weyl node and multiply the final result by $2N$.

In the presence of impurities and a neutralizing positively charged background, interacting electrons in the vicinity of a Weyl node are described by ($\hbar = 1$)
\begin{widetext}
\begin{equation}
\mathcal{L}=\Psi^\dagger(\mathbf{r},t) \left(i\partial_t + iv_f \mathbf{\sigma}\cdot\mathbf{\nabla} +\mu-V_{\text{dis}}(\mathbf{r})-V_{\text{ex}}(\mathbf{r})-ecA^0(\mathbf{r},t)\right)\Psi(\mathbf{r},t) - \frac{\epsilon c^2}{2} A^0(\mathbf{r},t) \nabla^2 A^0(\mathbf{r},t),
\end{equation}
\end{widetext}
where $\mathbf{\sigma} = (\sigma^1,\sigma^2,\sigma^3)$ refers to the Pauli matrices, $v_f$ is the Fermi velocity, $\mu$ is the chemical potential, $e$ is the fundamental electric charge, $c$ is the speed of light and $\epsilon$ is the dielectric constant. The fermion fields $\Psi^\dagger(\mathbf{r},t)$ and $\Psi(\mathbf{r},t)$ are 2-component Grassmann fields corresponding to creation and annihilation operators.  The random and static impurity potential $V_{\text{dis}}(\mathbf{r})$ assumes  white-noise correlator $\langle V_{\text{dis}}(\mathbf{r})V_{\text{dis}}(\pvec{r}')\rangle_{\text{dis}} \propto \delta(\mathbf{r}-\pvec{r})$ with zero mean $\langle V_{\text{dis}}(\mathbf{r}) \rangle_{\text{dis}} = 0$. We focus on ideal systems, almost free from impurities, so disorder is treated perturbatively to the lowest order in the fermion self-energy. The static potential energy from the jellium background reads as 
\begin{equation}
 V_{\text{ex}}(\mathbf{r}) = -n_0 \int d\pvec{r}'  V(\mathbf{r}-\mathbf{r}').
\end{equation}
Here $n_0$ denotes average density of the background ions identical to electron density in thermal equilibrium.

The $A^0$ field accounts for the Coulomb interaction between electrons. By integrating it out, one obtains the action for the Coulomb interaction reading as 
$S_C = -\frac{1}{2}\int d\mathbf{r}_1d\mathbf{r}_2dt \Psi^\dagger(\mathbf{r}_1,t)\Psi(\mathbf{r}_1,t) V(\mathbf{r}_1-\mathbf{r}_2) \Psi^\dagger(\mathbf{r}_2,t)\Psi^\dagger(\mathbf{r}_2,t).$
Here the Coulomb potential is $V(\mathbf{r}_1-\mathbf{r}_2) = \alpha v_f / |\mathbf{r}_1-\mathbf{r}_2|$, whose Fourier transform is given by $4 \pi \alpha v_f/q^2$, and $\alpha = e^2/(4\pi\epsilon v_f)$. The operator $-\nabla^2$ in the action for $A^0$ field should be understood as the inverse Fourier transform of $q^2$.

\begin{figure}[t]
    \begin{subfigure}{0.45\textwidth}
         \includegraphics[width=\textwidth]{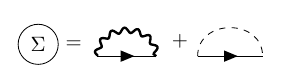}
        \caption{}\label{fig:fermionselfenergy}
    \end{subfigure}
     \begin{subfigure}{0.3\textwidth}
         \includegraphics[width=\textwidth]{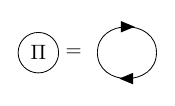}
         \caption{}
         \label{fig:plasmonselfenergy}
    \end{subfigure}
    \caption{Self-energy diagrams for fermion and boson propagators.  The thick wiggly line represents the dressed propagator $D(\mathbf{r},t;\mathbf{r}',t')$  of $\tilde{A}^0$. The dashed line denotes the disorder correlator $\langle V_{\text{dis}}(\mathbf{r}) V_{\text{dis}}(\mathbf{r}') \rangle$. The straight line is corresponding to the bare fermionic propagator $G_0(\mathbf{r},t;\mathbf{r}',t')$. }
\end{figure}

\subsection{Plasmon Dispersion}

Our analysis of the interaction effects begins with the decomposition of the $A^0$ field into a mean field and a quantum fluctuation. The saddle-point configuration denoted by $\bar{A}^0$  balances out the background potential $V_{\text{ex}}(\mathbf{r})$. This is in the spirit of the Hartree theory of interacting electrons \cite{Stoof2009,Kitinan2022b}. Fluctuations around this configuration are accounted for by the fluctuating field $\tilde{A}^0(\mathbf{r},t) = A^0(\mathbf{r},t) - \bar{A}^0$. Its propagator obeys a Dyson equation 
\begin{equation} 
D^{-1}(\mathbf{r},t;\mathbf{r}',t') = D^{-1}_0(\mathbf{r},t;\mathbf{r}',t') - \Pi(\mathbf{r},t;\mathbf{r}',t').
\label{eq:Dysonequationboson}
\end{equation}
Here $\Pi(\mathbf{r},t;\mathbf{r}',t')$ is the self-energy. For the purpose of this paper, it is sufficient to limit ourselves to the random phase approximation (RPA). We include only the bare polarization bubble diagram presented in Fig.\ref{fig:plasmonselfenergy} in the self-energy. The polarization function brings the plasmon poles to the dressed propagator. 

In non-equilibrium quantum field theory, transport properties are primarily determined by two main propagators, the so-called, retarded and Keldysh Green functions. The retarded Green function is denoted by $D^r(\mathbf{r},t;\mathbf{r}',t')$ for the bosonic field $\tilde{A}^0$. It obeys the Dyson equation (\ref{eq:Dysonequationboson}):
\begin{equation}\label{eq:dyson_equation_boson_retarded}
\left(D^{-1}\right)^{r}(\mathbf{r},t;\mathbf{r}',t') = \left(D_0^{-1}\right)^r(\mathbf{r},t;\mathbf{r}',t') - \Pi^r(\mathbf{r},t;\mathbf{r}',t')
\end{equation}

In momentum space, the bare propagator has no dynamics as the Coulomb interaction is instantaneous in the non-relativistic limit. It is given by $D_0^{-1}(\omega,\mathbf{q}) = q^2/4 \pi v_F \alpha$. Its dynamics, however, are generated by its interaction with electrons in terms of the self-energy $\Pi^r(\omega,\mathbf{q})$. The real part of the dressed propagator, $\Re D^r(\omega,\mathbf{q})$, shows poles at quasi-particle energy 
$\omega=\omega_{\text{pl}}(\mathbf{q})$. Here $\omega_{\text{pl}}(\mathbf{q})$ defines the plasmon energy dispersion. It reads

\begin{equation}\label{plasmon_dispersion}
\omega^2_{\text{pl}}(\mathbf{q}) =
     \frac{3}{5}v_f^2\mathbf{q}^2 + \frac{2Ne^2}{2\pi^2\epsilon \epsilon_r v_f}M^2(T,\mu),
\end{equation}
where the energy gap is given by $M^2(T,\mu) \equiv \frac{\pi^2}{3}T^2 + \mu^2$ \cite{le2000thermal,2013_Lv_Dielectric,2015_Hofman_Plasmon,2015_Kharzeev_Universality}. The permittivity is renormalised by the interaction and becomes $\epsilon_r = 1 + \frac{2Ne^2}{12\pi^2\epsilon v_f}\log\left(\frac{\Lambda}{\textrm{max}(T,\mu)}\right)$ \cite{Hosur2013}. Here $\Lambda$ is the energy bandwidth beyond which the band deviates from linear.

The imaginary part, $\Im D^r(\omega,\mathbf{q})$, gives the spectral density. It typically has peaks at the quasi-particle energy with broadening set by the decay rate $\gamma_{\text{pl}}$. In the long-wavelength limit it reads \cite{2015_Hofman_Plasmon,2015_Kharzeev_Universality}
\begin{gather}\nonumber
    \gamma_{\text{pl}} =  -\frac{\Im \Pi^r}{\partial_\omega \Re\Pi^r}\biggr\rvert_{\omega = \omega_\mathbf{q}} =
    \\
    \label{landau_damping}
    =  -\frac{\omega_{\text{pl}}^3}{8M^2}\left[\frac{1}{e^{\beta\left(\frac{\omega_{\text{pl}}}{2} - \mu\right)} + 1} - \frac{1}{e^{\beta\left(-\frac{\omega_{\text{pl}}}{2} - \mu\right)} + 1} \right].
\end{gather}

The quasi-particle approximation is valid as long as the spectral broadening is negligible. For the $\tilde{A}^0$ field, quasi-particle exist only in the low-momentum region below a characteristic momentum $q_c$ which is of the order of the inverse Thomas-Fermi screening length. Thus, the plasmons have as equal integrity as the other quasiparticles in the system, for example, as momentum and heat carriers.  For higher momenta, the quasi-particle approximation ceases to hold. Only a negligibly small contribution to momentum and heat fluxes can be expected from this incoherent mode. However, it still pays a crucial role as a mediator of electron-electron interaction. The interaction is effective at a shorter distance.

\subsection{Kinetic Equation}
The statistical properties of the plasmons are captured by their Keldysh Green's function
\begin{equation}
D^K(\omega,\mathbf{q}) = 2 i \Im D^r(\omega,\mathbf{q})(1+2b(\omega)).
\end{equation}
In thermal equilibrium the distribution function $b(\omega)$ reduces to the Bose-Einstein function evaluated at energy $\omega$. The spectral function $\Im D^r(\omega,\mathbf{q})$ describes the fuzziness  around the quasi-particle energy due to emission and absorption processes. In vacuum, the imaginary part is constrained by the mass-shell condition, i.e., $\Im G^r(\omega,\mathbf{q}) = -\pi \delta(\omega-\omega_{\text{pl}}(\mathbf{q})).$ The Keldysh Green's function within the lowest order in the gradient expansion obeys the Keldysh equation:
\begin{widetext}
\begin{equation}
	 \left(Z^{-1}(\mathbf{q},\omega)\partial_T + \mathbf{v}_b(\mathbf{q},\omega) \cdot \partial_{\mathbf{R}}  \right) D^K(\mathbf{q},\mathbf{R},\nu,T) \\ =  -2\Im\Pi^r(\mathbf{q},\omega)D^K(\mathbf{q},\omega)+2\Pi^K(\mathbf{q},\omega) \Im D^r(\mathbf{q},\omega),
\end{equation}
\end{widetext}
where the quasi-particle weight is given by $Z(\mathbf{q},\omega) = \partial_{\omega}(D^{-1}_0(\mathbf{q},\omega)- \Re \Pi(\mathbf{q},\omega))$ and the renormalized velocity  $\mathbf{v}_b(\mathbf{q},\omega) = \mathbf{\nabla}_{\mathbf{q}} \left(D^{-1}_0(\mathbf{q},\omega)-\Re \Pi(\mathbf{q},\omega)\right)$. As discussed in the previous subsection, in the low-momentum limit, plasmons behave as well defined quasi-particles. This allows to reduce the Keldysh equation to the Boltzmann equation:

\begin{widetext}
\begin{equation}
\left( \partial_t + \mathbf{v}_{\text{pl}}\cdot \mathbf{\nabla} \right) b(\mathbf{q},\mathbf{r},t) = \int_{\mathbf{p}} \bar{\mathcal{M}}_{\gamma\gamma'}(\mathbf{q},\mathbf{p}) \left[ (1-f_{\gamma}(\mathbf{p}))f_{\gamma'}(\mathbf{q}-\mathbf{p})b(\mathbf{q})-f_\gamma(\mathbf{p})(1-f_{\gamma'}(\mathbf{q}-\mathbf{p}))(1+b(\mathbf{q}))  \right],
\label{eq:plasmonboltzmann}
\end{equation}
\end{widetext}
where the plasmon's group velocity $\mathbf{v}_{\text{pl}} = \partial \omega_{\text{pl}}(\mathbf{q})/\partial \mathbf{q}$ and $f_\gamma$ is the quasiparticle distribution of the fermion species $\gamma$.

It should be mentioned, though, that close to zero temperature, the RPA gives a diverging scattering time as a result of the high stability of the plasmons against  Landau damping. Other competitive relaxation processes, such as, plasmon-impurity scattering, plasmon-plasmon interaction \cite{MARKOV2002} and non-linear plasmon-electron scattering, are required.  
 
\subsection{Fermion Green functions and kinetic equation}

The behaviour of the Weyl fermions is also captured by the dressed Green function which satisfies a Dyson eqaution:
\begin{equation} 
G^{-1}(\mathbf{r},t;\mathbf{r}',t') = G^{-1}_0(\mathbf{r},t;\mathbf{r}',t') - \Sigma(\mathbf{r},t;\mathbf{r}',t').
\label{eq:Dysonequationfermion}
\end{equation}

We assume that the Weyl fermions are well-defined quasi-particles. Their spectral function  is thus proportional to the Dirac delta function.

However, the Keldysh Green function which plays the role of the quantum counterpart of the phase-space distribution function is non-trivial and obey a Keldysh equation. To lowest order in the gradient expansion and the on-the-mass-shell condition, the Keldysh equation reduces to a Boltzmann equation:

\begin{widetext}
\begin{gather}
	 \left(\partial_t + \mathbf{v}_\gamma \cdot \mathbf{\nabla} - e\mathbf{E} \cdot \mathbf{\nabla}_{\mathbf{p}}\right)f_\gamma(\mathbf{p},\mathbf{r},t) = \int_{\mathbf{q}} \mathcal{M}_{\gamma\gamma'}(\mathbf{p},\mathbf{q}) \left[ (1-f_\gamma(\mathbf{p}))f_{\gamma'}(\mathbf{p}-\mathbf{q})b(\mathbf{q})-f_{\gamma}(\mathbf{p})(1-f_{\gamma'}(\mathbf{p}-\mathbf{q}))(1+b(\mathbf{q}))  \right] \nonumber\\ 
     +  \int_{\mathbf{q}} \tilde{\mathcal{M}}_{\gamma\gamma'}(\mathbf{p},\mathbf{q}) \left[ (1-f_\gamma(\mathbf{p}))f_{\gamma'}(\mathbf{p}+\mathbf{q})(1+b(\mathbf{q}))-f_{\gamma}(\mathbf{p})(1-f_{\gamma'}(\mathbf{p}+\mathbf{q}))b(\mathbf{q})  \right].
     \label{eq:fermionboltzmann}
\end{gather}
\end{widetext}
The coupled Boltzmann equations \eqref{eq:fermionboltzmann} and \eqref{eq:plasmonboltzmann} render a starting point for our calculation of transport coefficients.

\subsection{Effective Theory Approach}
In the previous section, the coupled system of Boltzmann equations (\ref{eq:plasmonboltzmann}) and(\ref{eq:fermionboltzmann}) for plasmons and fermions was derived. Of particular importance is the scattering amplitude $\mathcal{M}_{\gamma \gamma'}$ which depends on the absorption/emisison of plasmons by fermions. Here we will outline an effective theory approach to calculate this term. Besides being straightforward to apply in the case of plasmons, it can also be used for other electromagnetic collective excitations. Later in section \ref{discussion_section}, we will consider it for the case of helicons, gapless excitations present in a cold magnetic plasma (see e.g., Ref. \cite{alexandrov1984principles}).

First, we note that Weyl/Dirac fermions are coupled to internal electromagnetic (EM) degrees of freedom, represented by the 4-vector potential $A_\mu \equiv (A^0, - \mathbf{A})$, through the coupling $eA_\mu\Psi^\dagger \sigma^\mu\Psi$, where $\mu = 0,1,2,3$ and $\sigma^0 = I$. Hence, the effective Lagrangian for the electromagnetic degrees of freedom reads
\begin{gather}\label{matter_EM_Lagrangian}
    \mathcal{L}_A = -\frac{\epsilon c^2}{2}A^{\mu}_{-Q}\left(K_{\mu\nu} + \Re\Pi_{\mu\nu}\right)A^{\nu}_{Q} 
\end{gather}
where $A^\mu_Q$ is the electromagnetic 4-potential at 4-momentum $Q = (\omega, \mathbf{q})$. Furthermore, $K_{\mu\nu}$ is the vacuum kinetic matrix in the Coulomb gauge for the 4-potential, i.e.,
$K_{\mu\nu} = (\omega^2/c^2 - q^2)\eta_{\mu\nu} - \omega^2/c^2\delta_{\mu 0}\delta_{\nu 0}$, and $\Pi_{\mu\nu}$ is the polarization tensor \cite{le2000thermal}. Note that its 00-component is just the polarization $\Pi^r$  of the $A^0$ field considered in Eq. (\ref{eq:dyson_equation_boson_retarded}).
In Eq. (\ref{matter_EM_Lagrangian}) we have in principle kept all terms depending on both the scalar potential $A^0$ and the vector potential $\mathbf{A}$. However in the absence of a magnetic field in the Coulomb gauge, one has that $\Pi_{i0}A^i=\Pi_{0j}A^j = 0$ (in the case of Dirac fermions as the ones presented here, see e.g., Ref. \cite{le2000thermal}) and $A^0$ is decoupled from $\mathbf{A}$. Hence we will disregard the latter's contributions to Eq. (\ref{matter_EM_Lagrangian}) as is standard in condensed matter contexts, where only the internal longitudinal degree of freedom of the EM field is considered through the Coulomb interaction. 

The solutions $\omega = \omega_\mathbf{q}$, where the kinetic term is zero will be given by the condition $-q^2 + \Re\Pi_{00}(Q) = 0$ as for standard RPA plasmons. Expanding around this pole to quadratic order, one obtains the Lagrangian
\begin{gather}\nonumber
    \mathcal{L}_{p} = -\frac{\epsilon c^2 Z_{\mathbf{q}}^{-1}}{2}A_{-Q}^0[\omega^2 - \omega_\textrm{pl}^2(\mathbf{q})]A_Q^0 \\ 
    \label{effective_Lagrangian_plasmon}
    = \frac{1}{2}\varphi_{-Q}[\omega^2 - \omega_\textrm{pl}^2(\mathbf{q})]\varphi_Q ,
\end{gather}
where $Z_{\mathbf{q}}^{-1} \equiv \frac{\partial_\omega\Re\Pi_{00}|_{\omega=\omega_{\textrm{pl}}}}{2\omega_{\textrm{pl}}}$ is the renormalization factor. To get the canonical kinetic term for the plasmon, we have defined a new \lq\lq plasmon" field $\varphi_Q \equiv (\epsilon c^2Z_ {\mathbf{q}}^{-1})^{\frac{1}{2}}A^0_Q$ in Eq. (\ref{effective_Lagrangian_plasmon}). This change is consequential in more than just an arbitrary redefinition as it can be used to calculate the plasmon linewidth. 

In the conventional RPA approach, the plasmon linewidth is obtained by expanding $-c^2q^2 + \Pi_{00}(Q) = 0$ (notice that now we are not using only the real component of $\Pi_{00}$) around the pole $\omega = \omega_\mathbf{q} + i\gamma_\mathbf{q}$, assuming $\gamma_\mathbf{q}\ll \omega_\mathbf{q}$. This yields the decay linewidth given in Eq, (\ref{landau_damping}). This result can also be obtained by considering the electromagnetic field-fermion interaction,
\begin{equation}\label{plasmon_interaction}
 e c\int_P A^0_Q\Psi^\dagger_{P-Q}\Psi_P = \frac{eZ^{\frac{1}{2}}_ {\mathbf{q}}}{\sqrt{\epsilon}}\int_P \varphi_Q\Psi^\dagger_{P-Q}\Psi_P ,
\end{equation}
and Fermi's golden rule. Only the density-scalar potential interaction is considered as the interaction between the current and the vector potential will in comparison be suppressed by factors of $v_f/c$.  The square of the modulus of the scattering amplitude for plasmon absorption/emission between the fermionic species $\gamma$ and $\gamma'$ is thus given by $\mathcal{M}_{\gamma \gamma'} = \frac{e^2 Z_\mathbf{q}}{2\epsilon \omega_\mathbf{q}}\mathcal{F}_{\gamma\gamma'}(\mathbf{p},\mathbf{q}) = \frac{e^2}{\epsilon \partial_\omega \Re\Pi_{00}}\mathcal{F}_{\gamma\gamma'}(\mathbf{p},\mathbf{q})$, where $\mathcal{F}_{\gamma\gamma'}(\mathbf{p},\mathbf{q})$ is a structure factor, purely of fermionic origin. Fermi's golden rule then yields that the decay rate is just the linewidth calculated with the RPA approach,
\begin{gather}
    \gamma_\textrm{pl} \sim \sum_{\gamma,\gamma'}\int_\mathbf{p} \mathcal{M}_{\gamma \gamma '} \delta(\omega_\textrm{pl}(\mathbf{q}) + \varepsilon_{\gamma,\mathbf{p}} - \varepsilon_{\gamma,\mathbf{p} + \mathbf{q}}).
\end{gather}
Analogously, the interaction in Eq. (\ref{plasmon_interaction}) can also be used to straightforwardly derive the collision integral terms for the fermions in the Boltzmann equations in Eq. (\ref{eq:fermionboltzmann}).

One further detail that must be regarded is that plasmons are only well-defined excitations up to a certain energy/momentum. In our small momentum approximation, it is seen that $\gamma_\textrm{pl} \ll \omega_\textrm{pl} $ \cite{2015_Hofman_Plasmon,2015_Kharzeev_Universality}. However, for the momentum $q_c$, defined by $\omega_{\mathbf{q}_c} = v_f|\mathbf{q}_c|$, i.e., the limit of our approximation, one expects that plasmon decay to electron/hole pairs becomes an important source of decay. Therefore we take $\omega_{\mathbf{q}_c}$ as the cutoff for plasmon energies. This cutoff plays a role analogous to the Debye energy $\omega_\textrm{D}$ used to model acoustic phonons. As is known, in that case there is a different behavior between the $T\ll \omega_\textrm{D}$ and $T \gg \omega_\textrm{D}$, where in the former the leading small angle scattering gives an electric current relaxation time different from the naive lifetime estimate, see e.g. Ref. \cite{2019_Lavasani_WF}. However, while thermally activated plasmons relax electronic currents, by virtue of being neutral excitations they don't make a direct contribution to the electric current themselves, much like phonons in conventional transport theory.

\section{Transport}
\label{sec:transport}
\subsection{Thermoelectric Transport}
\begin{figure*}[t]
    \centering
    \begin{tabular}{cc}
        \includegraphics[width = 0.49\textwidth]{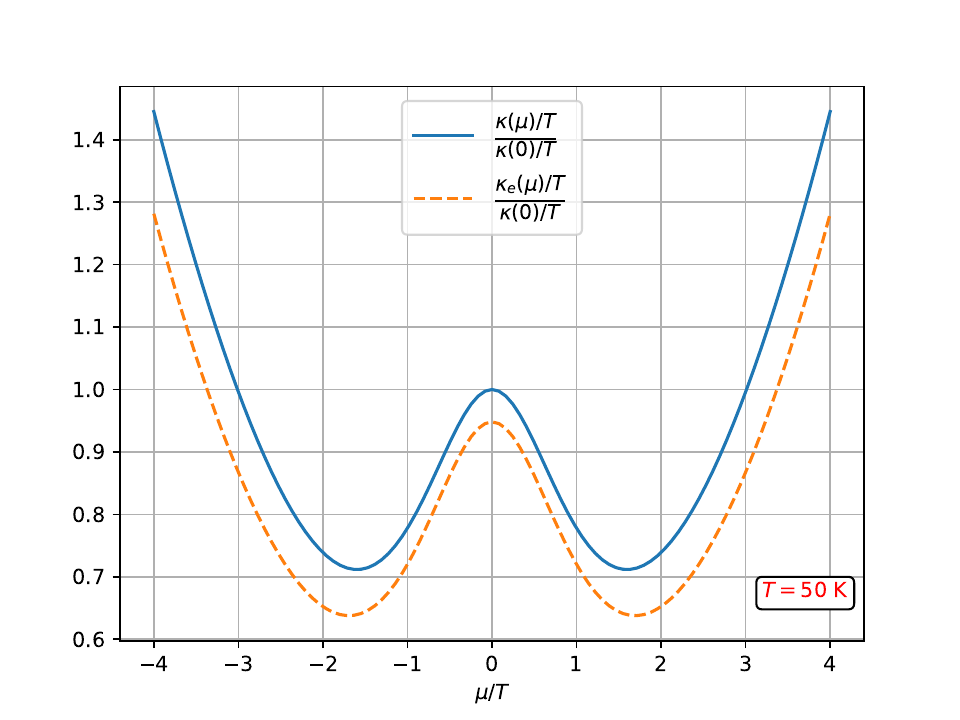} &  \includegraphics[width = 0.49\textwidth]{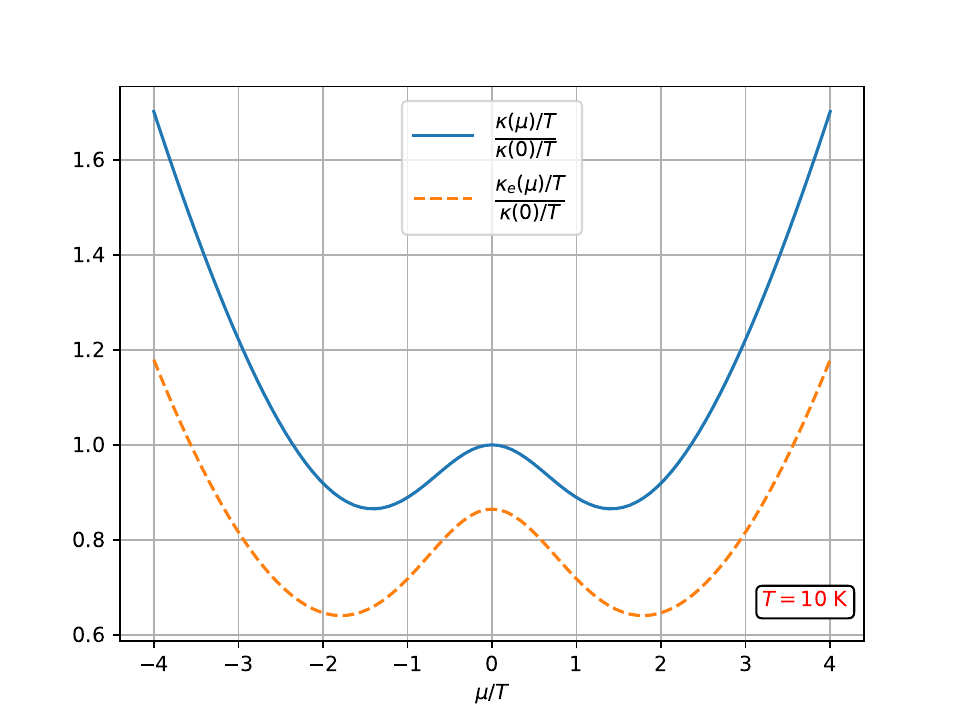} \\
         (a) & (b) \\ 
         \includegraphics[width = 0.49\textwidth]{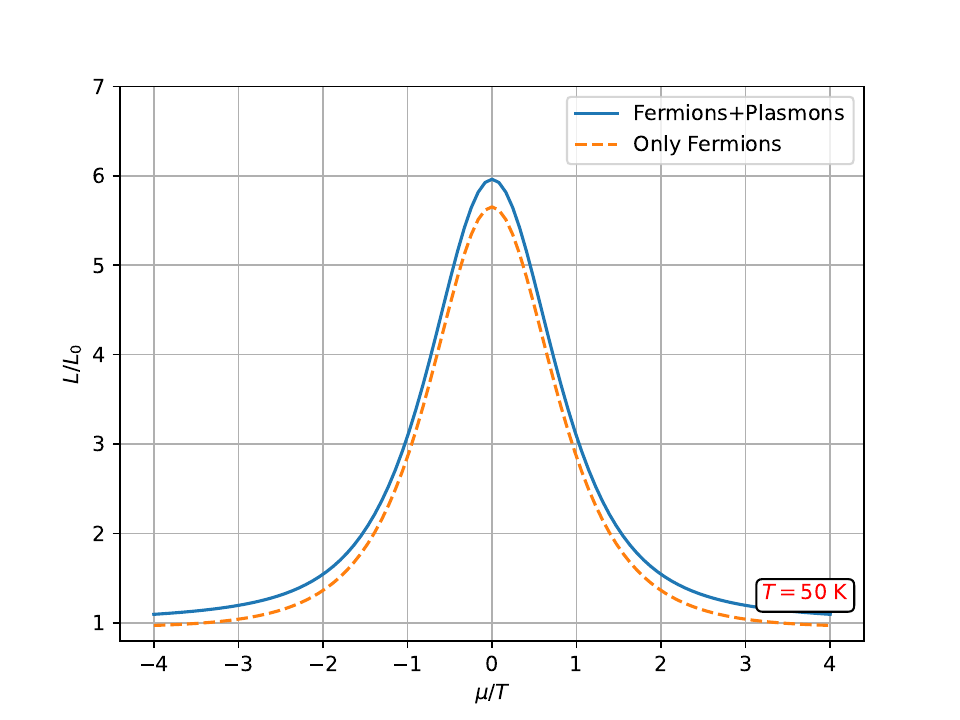}   &
         \includegraphics[width = 0.49\textwidth]{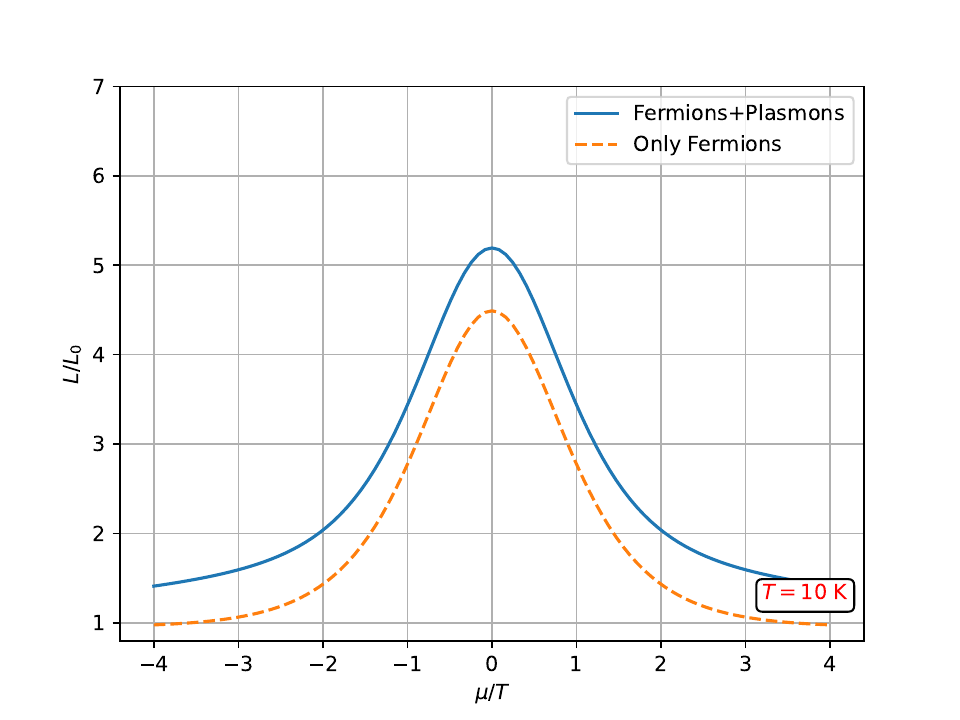}   \\
         (c) & (d)
    \end{tabular}

    \caption{(a-b) Thermal conductivities and (c-d) Lorenz Ratios at temperatures $T=50$ K and $T= 10$ K  for the fermion-plasmon system (blue) and the exclusively fermionic theory (dashed orange). We have taken $\alpha = 1/137$ and $c/v_f = 300$ and $\tau_{\textrm{imp}} = 3\cdot 10^{-13}$ s. }
    \label{fig:transport}
\end{figure*}
It is a well-known fact that the Boltmann equation, even for a single particle species, is generally impossible to solve exactly. Coupled Boltzmann equations further exacerbate this issue. Therefore, approximate solution schemes are generally implemented to obtain results. A common approach involves linearizing the equation by expanding $f_\gamma = f^0_\gamma + \delta f_\gamma$, where
\begin{gather}
    f^0_\pm \equiv f^0(\varepsilon_{\pm,\mathbf{p}}) = \frac{1}{e^{\beta(\varepsilon_{\pm,\mathbf{p}} - \mu)} + 1}
\end{gather}
is the Fermi-Dirac distribution for fermions in thermal equilibrium at temperature $T = \beta^{-1}$ (we take $k_\textrm{B} = 1$ so that temperature is measured in units of energy) and chemical potential $\mu$. In our case, the fermions can be electrons (+) and holes (-), which have dispersion $\varepsilon_{\pm,\mathbf{p}} = \pm v_f|\mathbf{p}|$. Additionally,
\begin{gather}
    f^0_b \equiv b^0(\varepsilon_{b,\mathbf{p}}) =  \frac{1}{e^{\beta\varepsilon_{b,\mathbf{p}}} - 1}
\end{gather}
is the Bose-Einstein distribution for bosons in thermal equilibrium. In the case at hand, these bosons are the plasmons with dispersion $\varepsilon_{b,\mathbf{p}}=\omega_\textrm{pl}(\mathbf{p})$ given in Eq. (\ref{plasmon_dispersion}). The linear perturbations $\delta f_\gamma$ are assumed to be of the order of the applied external perturbation $\mathbf{E}$, or $\nabla T$. Although linearization greatly decreases the complexity of the problem, further approximation schemes must be employed to solve the Boltzmann equation. 

An approach is to expand $\delta f$ in a physically-motivated basis of functions and convert the problem into a simple linear algebra problem \cite{ziman1960electrons,PeterArnold_2000}. Here we will adopt an alternative to this approach, consisting of a relaxation-time approximation (RTA) where the collision operator simply becomes a matrix in species space and whose elements have to be estimated through other means \cite{Kitinan2022a}, as in the standard single-species RTA. This latter method has been shown to reproduce the effects of the former in graphene in Ref. \cite{Kitinan2022c}, host of the 2D analogue of the Dirac fermions found in WSM.

After linearizing and implementing the RTA described above, the coupled Boltzmann equations  in the presence of an external electric field and thermal gradient become
\begin{widetext}
    \begin{align}
    - (\varepsilon_{+,\mathbf{p}} - \mu)\beta\nabla T \cdot \partial_{\mathbf{p}}f^0_+ - e\mathbf{E}\cdot \partial_{\mathbf{p}}f^0_+ &= -\delta f_+\left(\frac{1}{\tau_{+-}} + \frac{1}{\tau_\textrm{+imp}} + \frac{1}{\tau_{+b}}\right) + \frac{\delta f_-}{\tau_{-+}} + \frac{\delta f_b}{\tau_{b+}},  \label{BE_linearized_thermo_electric_electrons} \\
    - (\varepsilon_{-,\mathbf{p}} - \mu)\beta\nabla T \cdot \partial_{\mathbf{p}}f^0_- - e\mathbf{E}\cdot \partial_{\mathbf{p}}f^0_- &= \frac{\delta f_+}{\tau_{+-}} -\delta f_-\left(\frac{1}{\tau_{-+}} + \frac{1}{\tau_\textrm{-imp}} + \frac{1}{\tau_{-b}}\right)  + \frac{\delta f_b}{\tau_{b-}}, \label{BE_linearized_thermo_electric_holes}\\
     - \varepsilon_{b,\mathbf{p}}\beta\nabla T \cdot \partial_{\mathbf{p}}f^0_b  &= \frac{\delta f_+}{\tau_{+b}} + \frac{\delta f_-}{\tau_{-b}}  - \delta f_b\left(\frac{1}{\tau_{b+}} + \frac{1}{\tau_{b-}}\right) .
     \label{BE_linearized_thermo_electric_plasmons}
    \end{align}
\end{widetext}
The relaxation times are parametrized in such a way that, in the absence of a coupling of the fermions to bosons and impurities, fermion number, momentum and energy currents are all conserved, as can be readily checked \cite{Kitinan2022a}.

Separating the Boltzmann equations (\ref{BE_linearized_thermo_electric_electrons}-\ref{BE_linearized_thermo_electric_plasmons}) into electric and thermal components, they can be recast into the form $S^q_\gamma = - C_{\gamma \gamma'}\delta f^q_{\gamma'} $ where the index $q$ indicates transport by an external electric field ($q = E$) or a thermal gradient ($q=T$) and $\gamma,\gamma'=\pm,b$ are particle indices as before. The values of the source terms $S^q_{\gamma \mathbf{p}}$ are given in Table \ref{linear_BE_table}, while the collision matrix is given by
\begin{gather}\nonumber
C= \\ 
    \begin{pmatrix}
       \scriptstyle \frac{1}{\tau_{+-}} + \frac{1}{\tau_{\textrm{imp}}} + \frac{1}{\tau_{+b}}&& \scriptstyle - \frac{1}{\tau_{-+}}  && \scriptstyle -\frac{1}{\tau_{b+}} \\ \scriptstyle
        - \frac{1}{\tau_{+-}}  && \scriptstyle  \frac{1}{\tau_{-+}} + \frac{1}{\tau_{\textrm{imp}}} + \frac{1}{\tau_{-b}}  && \scriptstyle \frac{1}{\tau_{b}} \\ 
        \label{collision_matrix_thermoelectric}
       \scriptstyle \frac{1}{\tau_{+b}} && \scriptstyle \frac{1}{\tau_{-b}}  && \scriptstyle \frac{1}{\tau_{+b}}+\frac{1}{\tau_{-b}}
    \end{pmatrix}.
\end{gather}
Defining $j^{q}$ as the charge ($q=E$) and energy ($q=T$) current, one finds
\begin{equation}
    \mathbf{j}^q \equiv \sum_{\gamma,\ q'} N_\gamma\int_{\mathbf{p}} Q^q_{\gamma}\mathbf{v}_{\gamma,\mathbf{p}} \delta f_\gamma^{q'} = \sum_{q'}\Sigma^{qq'}\mathbf{F}^{q'},
\end{equation}
where the conductivities are given by the trace of the matrix product
\begin{gather}
     \Sigma^{qq'} = \frac{1}{3}\textrm{tr}[C^{-1}\mathcal{T}^{qq'} ], \\
     \mathcal{T}^{qq'}_ {\gamma'\gamma} = N_{\gamma}\int_{\mathbf{p}}  \left(-\partial_\varepsilon f_{\gamma'}^0\right)Q^{q}_{\gamma,\mathbf{p}}Q^{q'}_{\gamma',\mathbf{p}}\mathbf{v}_{\gamma,\mathbf{p}}\cdot\mathbf{v}_{\gamma',\mathbf{p}}.
\end{gather}
Here, $N_\gamma$ is the number of $\gamma$ species, i.e., $N_+ = N_- = 2N$ and $N_b = 1$. In addition $\mathbf{F}^q$ are the applied generalized forces and $Q^q_\gamma$ are the electric/thermal charges of species $\gamma$. The values of these are also listed in Table \ref{linear_BE_table}. Furthermore, $\partial_\varepsilon f_{\gamma}^0$ should be understood as $\partial_\varepsilon f^0|_{\varepsilon=\varepsilon_{\gamma,\mathbf{p}}}$ for fermions and $\partial_\varepsilon b^0|_{\varepsilon=\varepsilon_{\gamma,\mathbf{p}}}$ for bosons.

\begin{table}[h!]
\begin{center}
\begin{tabular}{ |c |c |c |}
\hline 
  & $q = E$ & $q = T$\\
  \hline
 $Q_\pm^q$ & $-e$ & $(\epsilon_{\pm\mathbf{p}} - \mu)$\\
 $Q_b^q$ & $0$ & $\omega_{pl}(\mathbf{p})$\\
 \hline
 $S^{q}_{\pm\mathbf{p}}$ & $-e\nabla_\mathbf{p}f_{\pm\mathbf{p}}^0\cdot\mathbf{E} = $ & $-\beta(\epsilon_{\pm\mathbf{p}} - \mu)\nabla_\mathbf{p}f_{\pm\mathbf{p}}^0\cdot \nabla T $ \\
 $S^{q}_{b\mathbf{p}}$ & 0 &$-\beta\omega_{\mathbf{p}}\nabla_\mathbf{p}b_{\mathbf{p}}^0\cdot \nabla T$ \\
 \hline
 $\mathbf{F}^q$ & $\mathbf{E}$ & $-\beta\nabla T$ \\
 \hline
\end{tabular}
\caption{Values of transport charges $Q_\gamma^q$, source terms $S^{q}_{\gamma\mathbf{p}}$ and driving forces $\mathbf{F}^q$ for different species types and $q$ indices.}
\label{linear_BE_table}
\end{center}
\end{table}

It is interesting to study the case where $\mu = 0$, as then the plasmon reaches its smallest gap relative to temperature $\omega_\textrm{pl}(0)$ and leads to simplifications in the relaxation-time parameters due to the electron-hole symmetry $\tau_{+-} = \tau_{+-}$, $\tau_\textrm{+imp} = \tau_\textrm{-imp}\equiv \tau_\textrm{imp}$ , $\tau_\textrm{+b} = \tau_\textrm{-b}$, and $\tau_\textrm{b+} = \tau_\textrm{b-}$. Additionally, the electric conductivity reads
\begin{gather}
    \sigma(\mu = 0) = \Sigma^{EE}(\mu = 0)= \frac{4Ne^2\tau_\textrm{eff}}{3}\int_{\mathbf{p}}\left(-\partial_\varepsilon f_{+}^0\right)\mathbf{v}_{+\mathbf{p}}^2,  
\end{gather}
where $\tau_\textrm{eff} \equiv \left(\tau_{\textrm{imp}}^{-1} + \tau_{\textrm{+b}}^{-1} + 2\tau_{\textrm{+-}}^{-1}   \right)^{-1}$. The non-short-circuited thermal conductivity has the form
\begin{gather}\nonumber
    \bar{\kappa}(\mu = 0) = \beta\Sigma^{TT}(\mu = 0)= \\
    \nonumber
    \frac{\beta\tau_{\textrm{imp}}}{3}\int_{\mathbf{p}}\left(-4N\varepsilon_{+,\mathbf{p}}\mathbf{v}_{+,\mathbf{p}}\partial_\varepsilon f_{+}^0-\varepsilon_{b\mathbf{p}}\mathbf{v}_{b\mathbf{p}}\partial_\varepsilon f_{b}^0\right)
    \\ \nonumber
    \cdot\left(\varepsilon_{+,\mathbf{p}}\mathbf{v}_{+,\mathbf{p}} + \frac{\tau_{b+}}{2\tau_{+b}} \varepsilon_{b,\mathbf{p}}\mathbf{v}_{b,\mathbf{p}}\right) \\
    + \frac{\beta}{3}\frac{\tau_{b+}}{2}\int_\mathbf{p}\left(-\partial_\varepsilon f_{b}^0\right)\varepsilon_{b,\mathbf{p}}^2\mathbf{v}_{b,\mathbf{p}}^2.
\end{gather}

In systems where time-reversal symmetry is conserved, Onsager's reciprocal relations must be satisfied, i.e. $\Sigma^{ET} = \Sigma^{TE}$ must hold. In our model, this implies that constraints must be imposed on the collision matrix elements. After some simplifications (see Appendix \ref{appendix_collision_matrix}) a constraint relating $\tau_+$ and $\tau_-$ can be derived after imposing $\Sigma^{ET} = \Sigma^{TE}$. This constraint can be solved by parametrizing these relaxation times in terms of of another scale $\tau_0$, which we consider to be the plasmon Landau damping in Eq. (\ref{landau_damping}).

\begin{figure*}[t]
    \centering
    \begin{tabular}{cc}
        \includegraphics[width = 0.49\textwidth]{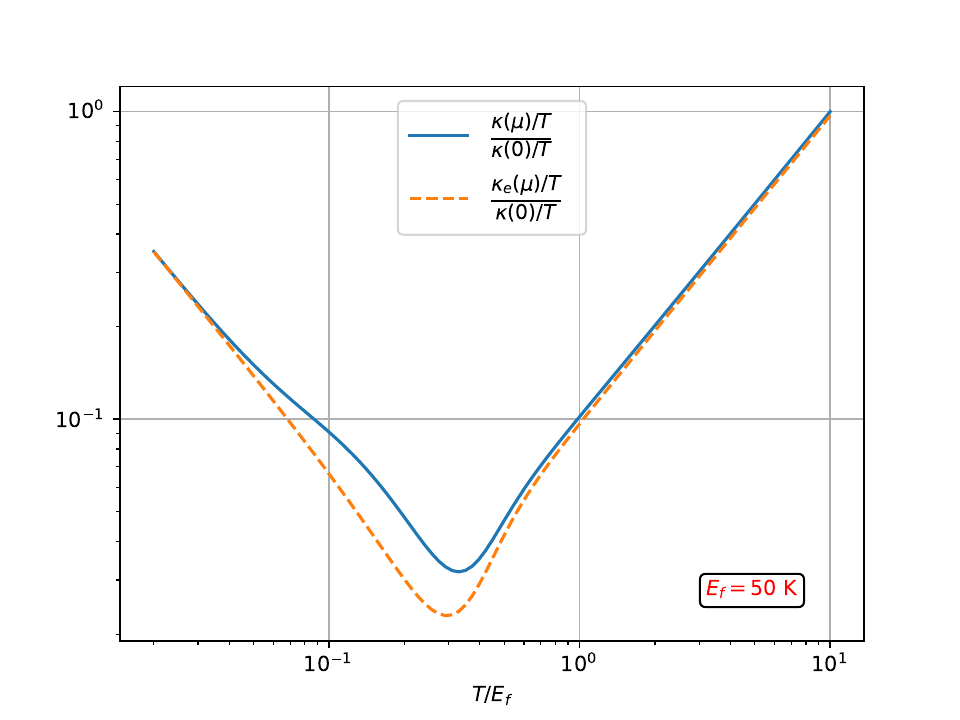} &  \includegraphics[width = 0.49\textwidth]{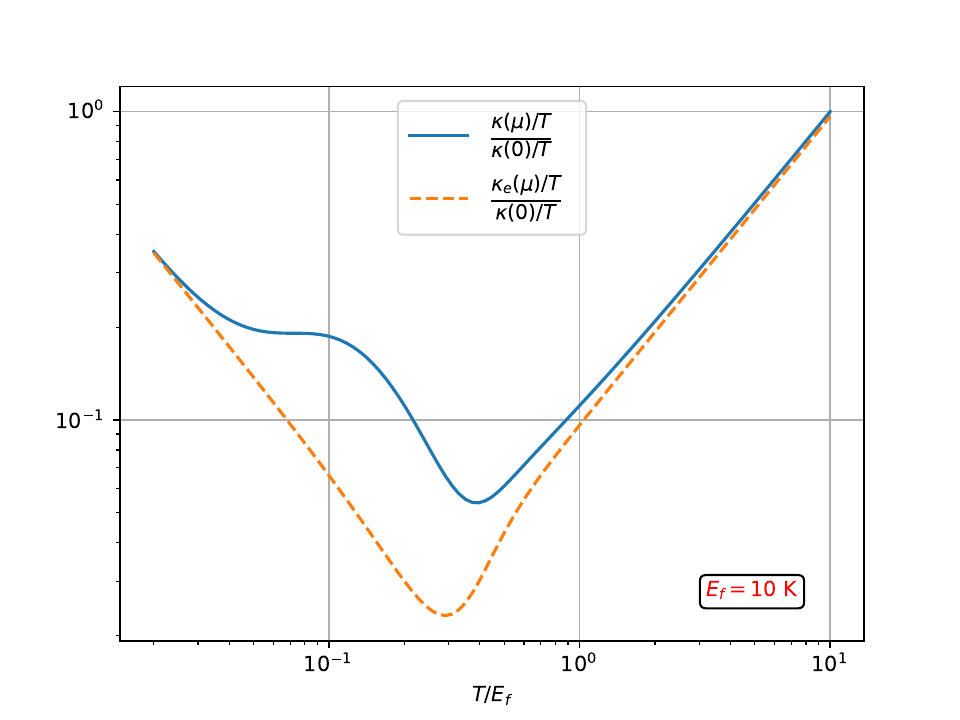} \\
         (a) & (b) \\ 
         \includegraphics[width = 0.49\textwidth]{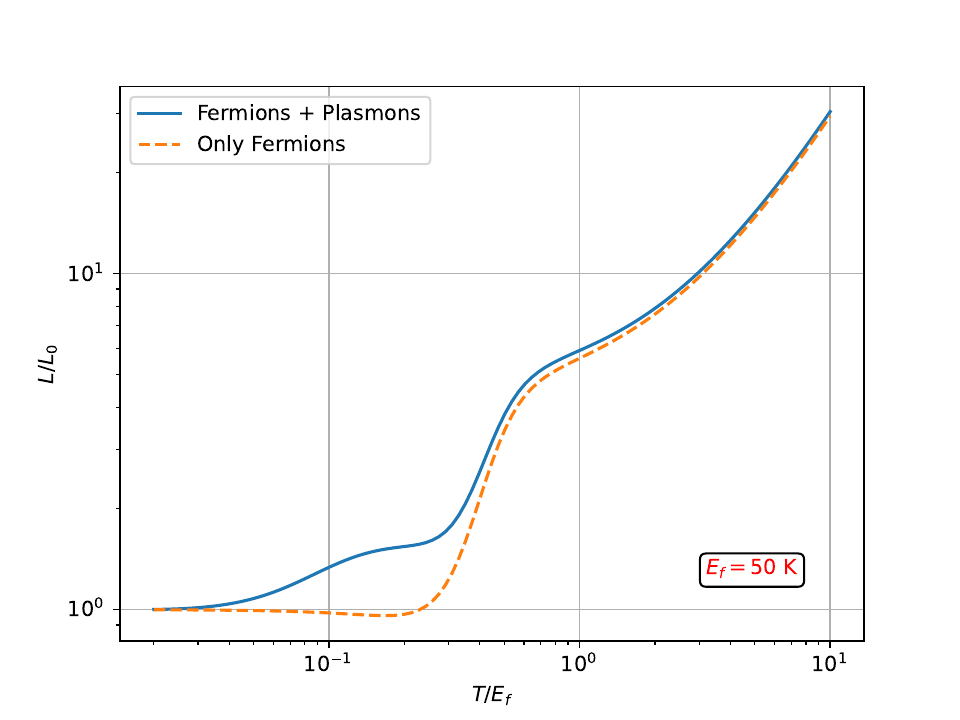}   &
         \includegraphics[width = 0.49\textwidth]{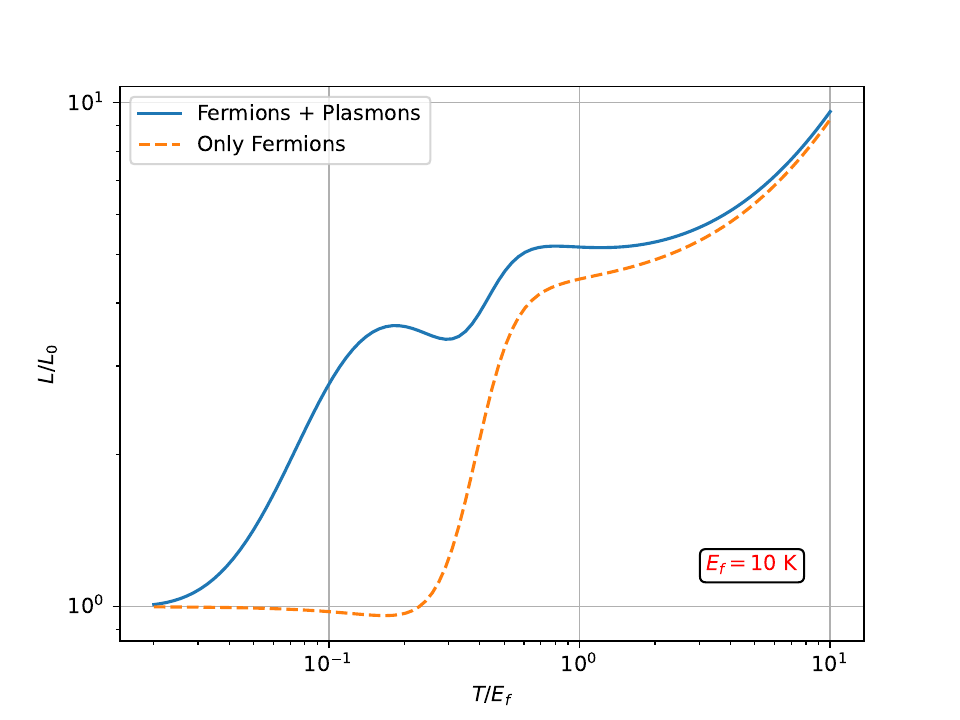}   \\
         (c) & (d)
    \end{tabular}

    \caption{(a-b) Thermal conductivities and (c-d) Lorenz Ratios at two different carrier densities for varying temperatures. The carrier densities are fixed by a Fermi Energy $E_f$. For (a) and (c) $E_f= 50 K$, whereas for (b) and (d) $E_f = 10 K$. Note that $E_f = \mu(T=0)$ for fixed carrier density. The thermal conductivities are referenced with respect to the value of the fermion and plasmon system at $T = 10E_f$. We have taken $\alpha = 1/137$ and $v_f = 10^6 m/s$ and $\tau_{\textrm{imp}} = 3\cdot 10^{-13}$ s.  }
    \label{fig:transport_fixed_Ef}
\end{figure*}

The resulting thermal conductivities and Lorenz ratio for the exclusively fermionic and the fermion-plasmon system can be seen in Fig. \ref{fig:transport} at $T=50$ K and an impurity scattering time of $\tau_\textrm{imp} = 0.3$ ps. Expectedly, the most notable effect of the plasmons is seen in the thermal transport, and hence the Lorenz ratio, as is the case in graphene \cite{Kitinan2022b}. Furthermore, this effect becomes more pronounced for small temperatures, where impurity scattering starts dominating the electron sector, see Fig. \ref{fig:transport}b and d for the case at $T=10 $K. For sufficiently large values of $|\mu/T|$ beyond those displayed in Fig.\ref{fig:transport}, the curves for the transport coefficients in the fermion-plasmon system would converge to the exclusively fermionic system as the plasmon gap in Eq. (\ref{plasmon_dispersion}) would be too large to be compensated by thermal fluctuations. At fixed carrier density (or equivalently, Fermi energy $E_f$), it is also seen that small temperatures offer the most distinguishable deviation of the thermal conductivity from a system devoid of plasmons, as shown in Fig. \ref{fig:transport_fixed_Ef}.

\subsection{Viscosity}
\begin{figure}[ht]
    \centering
    \includegraphics[width = 0.49\textwidth]{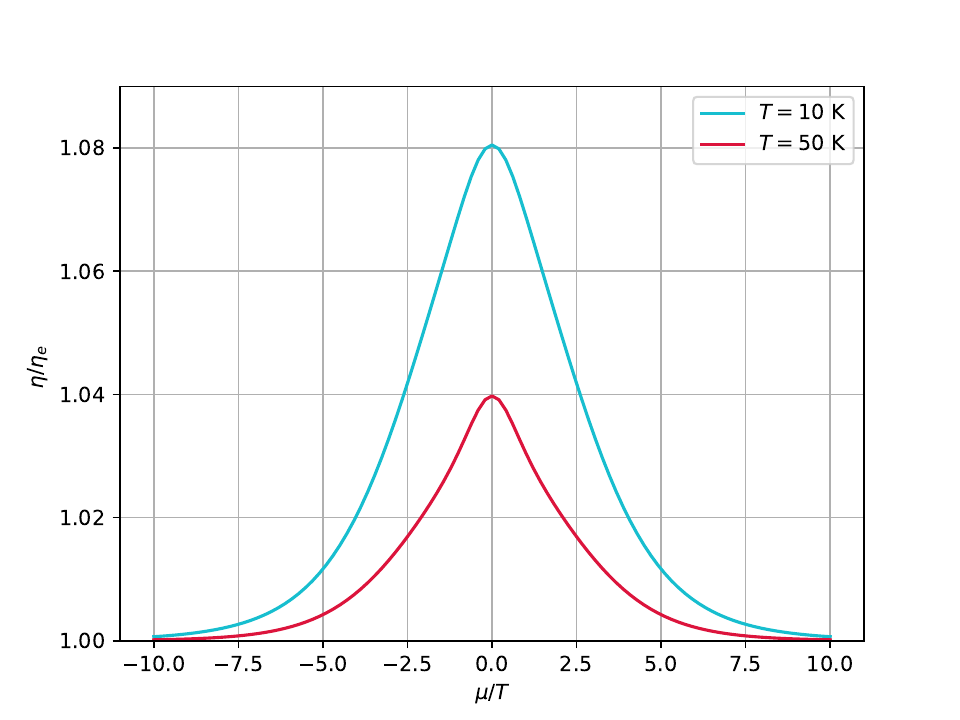}

    \caption{ Ratio between the viscosity in the fermion-plasmon system $\eta$ and the viscosity for the purely fermionic system $\eta_e$ for $T=50 $K and $T = 10$. The remaining parameters are the same as Fig. \ref{fig:transport}}
    \label{fig:viscosity_ratios}
\end{figure}

We can use a similar procedure as in the thermoelectric transport coefficients to calculate the shear viscosity, which is defined by $\pi_{ij} = \eta_{ijkl} X_{kl}$, where $\pi_{ij}$ is the momentum current tensor and $X_{kl} = \frac{1}{2}(\partial_{i}u_j + \partial_{j}u_i)$ describes small perturbations about the velocity $\mathbf{u}$ of the fermion-boson fluid and
\begin{equation}
    \pi_{ij} = \sum_{\gamma}N_\gamma \int_{\mathbf{p}} Q^\eta_{\gamma,\mathbf{p}} \hat{\mathbf{p}} \mathbf{v}_{\gamma, \mathbf{p}} \delta f_\gamma,
\end{equation}
where $Q^{\eta}_{\gamma,\mathbf{p}} = p$ is the charge associated to the viscosity, which is the same for fermions and bosons. Following a similar procedure as before, one finds that $\eta_{ijkl} = \eta\left[\delta_{ik}\delta_{jl}+\delta_{ik}\delta_{jl}-\frac{2}{3}\delta_{ij}\delta_{kl}\right]$, where
\begin{equation}\label{viscosity}
    \eta = \frac{1}{30}N_{\gamma'}C^{-1}_{\gamma \gamma'}\int_\mathbf{p} \left(-\partial_\epsilon f_{\gamma}^0\right) Q^{\eta}_{\gamma,\mathbf{p}} Q^{\eta}_{\gamma',\mathbf{p}}\mathbf{v}_{\gamma, \mathbf{p}}\cdot\mathbf{v}_{\gamma', \mathbf{p}}.
\end{equation} 
The ratios between the viscosity in the fermion-plasmon system and the purely fermionic case are shown in Fig.\ref{fig:viscosity_ratios} for two different temperatures. As was the case of the thermal conductivity, the plasmonic effects are higher at lower temperatures.

\section{Conclusion and outlook}\label{discussion_section}
In this paper we set up a full `phenomenological' hydrodynamic description of interacting Weyl systems in the framework of the Boltzmann equation. Experimentally, gaps of around 70 meV at room temperature have been observed in Weyl SMs such as TaAs and NbAs \cite{Chiarello2019}, and tuning of the doping and decrease in temperature should further contribute in lowering this gap, as seen in Eq. (\ref{plasmon_dispersion}). Such small gaps in the plasmon spectrum in comparison with with those found in conventional 3D metals, which are of the order of  $15-20$ eV, is what enables the thermal plasmon contribution to thermal conductivity and shear viscosity that we have demonstrated in this work. Indeed, using a simplified relaxation time approximation approach, we found a considerable increase in  thermal and viscosity coefficients under experimentally relevant conditions.

While this work is centered around plasmon effects, the microscopic theory in Eq. (\ref{matter_EM_Lagrangian}) and the Boltzmann equations (\ref{BE_linearized_thermo_electric_electrons}-\ref{BE_linearized_thermo_electric_plasmons}) could in principle be used to study interaction effects for other plasma collective modes. For transport effects as those considered in this work, it would be ideal to study fermion interactions with gapless, and hence thermally accessible, bosonic modes, as is the case of plasmons in graphene \cite{Kitinan2022a}. Helicons \cite{alexandrov1984principles,2015_Pellegrino_Helicons}, mixed longitudinal-transverse modes appearing in magnetized cold plasmas in a weak magnetic field $B$, i.e., $\omega_c \equiv |eB|/\mu \ll \mu$, satisfy this condition. In Weyl semimetals their dispersion is $\omega_{h,\mathbf{q}} = \frac{q|q_3|}{2M_h}$, where $2M_h \equiv \frac{\omega_p^2 + 2\alpha cb_3/\pi }{\omega_c}$, where $b_3$ is projection of the separation between the left and right Weyl nodes onto the direction of the applied magnetic field. It can be shown, (see Appendix \ref{appendix_helicons}) that for the helicon field $h$, the scalar potential-density interaction becomes
\begin{gather}\label{helicon_lagrangian}
     e c\int_P A^0_Q\Psi^\dagger_{P-Q}\Psi_P =eG^{\frac{1}{2}}_ {\mathbf{q}}\int_P h_Q\Psi^\dagger_{P-Q}\Psi_P,
\end{gather}
where
\begin{equation}\label{helicon_fermion_coupling}
    G^{\frac{1}{2}}_{\mathbf{q}} = \frac{\omega_{h,\mathbf{q}}^2}{c|q_3|\omega_p}.
\end{equation}
 Note that in spite of the fact that for $q = |q_3|$, $A^0_Q = 0$ so that the helicons are purely transverse, the interaction between the helicon modes and fermions is actually finite. However, due to the limited phase space for these excitations, scattering processes mediated by the interaction (\ref{helicon_fermion_coupling}) are strongly suppressed. Thus, it is not expected for helicons to contribute to electron transport.

In contrast, here it has been shown that despite being gapped, plasmons in nearly charge-neutral Weyl semimetals enhance transport signatures in a hydrodynamic regime, particularly in the relevant low temperature scenario. As was the case of graphene, these effects will be most significant in thermal and viscous effects.In contrast to the gapless plasmons of graphene, these effects will wane and become insignificant in a Fermi liquid phase, where thermal fluctuations will no longer be able to breach the plasmon frequency gap and excite plasmon fluctuations. 

\section*{Acknowledgments} 
We thank A. Cortijo, T. Ludwig, S. Grubinskas, P. Cosme, E. Di Salvo, T. Meng for prevoius collaborations and valuable discussions. J.B. is supported by the  FPU grant FPU20/01087. The research stay in Utrecht through which this project was conceived was funded by the Spanish Ministerio de Ciencia e Innovación grant PID2021-127240NB-I00. KP acknowledges financial support by the Deutsche Forschungsgemeinschaft (DFG) via the Emmy Noether Programme (Quantum Design grant, ME4844/1, project- id 327807255), project A04 of the Collaborative Research Center SFB 1143 (project-id 247310070), and the Cluster of Excellence on Complexity and Topology in Quantum Matter ct.qmat (EXC 2147, project-id 390858490).

\appendix
\section{Polarization Function}
To calculate the plasmon dispersion and Landau damping, we use the RPA approximation. The polarization function is given by $\Pi(\omega,\mathbf{q}) \equiv \sum_{\gamma \gamma'} \Pi_{\gamma \gamma '}(\omega,\mathbf{q})$
\begin{gather}\nonumber
    \Pi_{\gamma \gamma'}(\omega,\mathbf{q}) = 2N\int_{\mathbf{p}}\mathcal{F_{\gamma\gamma'}(\mathbf{p},\mathbf{p} + \mathbf{q}}) \\ 
    \cdot \frac{f^0(\epsilon_{\gamma\mathbf{p}}) - f^0(\epsilon_{\gamma'\mathbf{p} + \mathbf{q}})}{\omega + i\varepsilon + \epsilon_{\gamma\mathbf{p}} - \epsilon_{\gamma'\mathbf{p} + \mathbf{q}}},
\end{gather}
where
\begin{equation}
    \mathcal{F_{\gamma\gamma'}(\mathbf{p},\mathbf{p}'}) = \frac{1}{2}\left[1 + \gamma \gamma'\frac{\mathbf{p}\cdot\mathbf{p}'}{|\mathbf{p}||\mathbf{p}'|}\right].
\end{equation}
Assuming that $\omega, v_fq \ll \textrm{max}(\mu,T)$, we have
\begin{widetext}
    \begin{gather}\nonumber
    \Pi_{++} + \Pi_{--} \approx \frac{2N}{(4\pi^2)}\left[\int_{-1}^1 \frac{v_f qx dx}{\omega + i\varepsilon - v_f qx}\right]\left[-\int_0^\infty p^2 \frac{df^0(p,\mu)}{dp} dp\right] \\
    \nonumber
    +\frac{2N}{(4\pi)^2}\left[\int_{-1}^1 \frac{v_f qx dx}{\omega + i\varepsilon + v_f qx}\right]\left[\int_0^\infty p^2 \frac{df^0(p,-\mu)}{dp} dp\right] \\
    \nonumber
    = -2N\frac{T^2}{2\pi^2}\left[1 + \frac{\omega + i\varepsilon}{2v_f q}\log\left(\frac{\omega + i\varepsilon - v_f q}{\omega + i\varepsilon + v_f q}\right)\right]\left[-2\textrm{Li}_2(-e^{\beta\mu})\right] \\
    \nonumber
    - 2N\frac{T^2}{2\pi^2}\left[1 - \frac{\omega + i\varepsilon}{2v_f q}\log\left(\frac{\omega + i\varepsilon + v_f q}{\omega + i\varepsilon - v_f q}\right)\right]\left[-2\textrm{Li}_2(-e^{-\beta\mu})\right] \\
    \nonumber
    =-2N\frac{T^2}{2\pi^2}\left[1 - \frac{\omega + i\varepsilon}{2v_f q}\log\left(\frac{\omega + i\varepsilon + v_f q}{\omega + i\varepsilon - v_f q}\right)\right]\left[-2\textrm{Li}_2(-e^{\beta\mu})-2\textrm{Li}_2(-e^{-\beta\mu})\right] \\
    = -\frac{2N}{2\pi^2}\left[1 - \frac{\omega + i\varepsilon}{2v_f q}\log\left(\frac{\omega + i\varepsilon + v_f q}{\omega + i\varepsilon - v_f q}\right)\right]\left[\frac{\pi^2}{3}T^2 + \mu^2\right]. 
\end{gather}
The cross terms give
\begin{gather}
     \Pi_{+-} + \Pi_{-+} \approx \frac{2N}{3\pi^2} \frac{v_f^2q^2}{4}\int_0^\infty dp \left[\frac{1}{\omega + i\varepsilon + 2p}-\frac{1}{\omega + i\varepsilon - 2p}\right]\left[f^0(p,\mu) + f^0(p,-\mu) - 1\right],
\end{gather}
which can be decomposed into a real and an imaginary part as
\begin{gather}
     \Re[\Pi_{+-} + \Pi_{-+}] \approx -\frac{2N}{3\pi^2} \frac{v_f^2q^2}{4}\log\left(\frac{\Lambda}{\textrm{max}(T,\mu)}\right), \quad \Im[\Pi_{+-} + \Pi_{-+}] = -\frac{2Nv_f^2q^2}{24\pi}\left[\frac{1}{e^{\beta\left(\frac{\omega}{2} - \mu\right)} + 1} - \frac{1}{e^{\beta\left(-\frac{\omega}{2}- \mu\right)} + 1} \right].
\end{gather}
\end{widetext}
The plasmon is found from the solutions of
\begin{equation}
    \epsilon_\textrm{RPA}(\omega, \mathbf{q}) = 1 - V(\mathbf{q})\Pi(\omega, \mathbf{q}) = 0.
\end{equation}
Assuming solutions of the form $\omega = \omega_{\mathbf{q}} + i \gamma_{\mathbf{q}}$, this implies, the dispersion and damping are given respectively by
\begin{gather}
\label{appendix_plasmon_dispersion_condition}
    1 - V(\mathbf{q})\Re\Pi(\omega_\mathbf{q}, \mathbf{q}) = 0, \\
    \label{appendix_landau_damping_definition}
    \gamma_{\mathbf{q}} = -\frac{\Im\Pi(\omega_p,\mathbf{q})}{\partial_\omega\Re\Pi(\omega_p,\mathbf{q})}.
\end{gather}
Solving Eq. (\ref{appendix_plasmon_dispersion_condition}) in the $\omega \gg v_f q$ limit, one deduces Eq. (\ref{plasmon_dispersion}). Similarly, solving Eq. (\ref{appendix_landau_damping_definition}) gives Eq. (\ref{landau_damping}).

\section{Collision Matrix Elements}\label{appendix_collision_matrix}
\subsection{Thermoelectric Transport}
As was mentioned in the main text, Onsager reciprocity is not automatically guaranteed starting from the RTA approach of Eqs. (\ref{BE_linearized_thermo_electric_electrons}-\ref{BE_linearized_thermo_electric_plasmons}). Imposing $\Sigma^{ET} = \Sigma^{TE}$ leads to a convoluted constraint between the different relaxation times appearing in the problem. However, taking $\tau_{\pm b}^{-1} = 0$, i.e., a collision matrix of the form
\begin{gather} \label{collision_integral_approx}
C=
    \begin{pmatrix}
        \frac{1}{\tau_{+}} + \frac{1}{\tau_{\textrm{imp}}} && - \frac{1}{\tau_{-}}  && -\frac{1}{\tau_{b}} \\
        - \frac{1}{\tau_{+}}  &&  \frac{1}{\tau_{-}} + \frac{1}{\tau_{\textrm{imp}}}  && \frac{1}{\tau_{b}} \\ 
       0 && 0  && \frac{2}{\tau_{b}} 
    \end{pmatrix}
\end{gather}
significantly simplifies the problem, as now the Onsager condition reduces to a constraint exclusively for $\tau_{\pm}$ of the form
\begin{gather}\label{constraint}
    \sum_{s=\pm}\frac{1}{\tau_s}\left[\mathcal{A}_s +2\mathcal{B}_s + 2\mu\mathcal{C}_s \right] = 0,
\end{gather}
where
\begin{gather}
    \mathcal{A}_s = \int_{\mathbf{p}} (-\partial_\varepsilon f^0_b)\varepsilon_{b,\mathbf{p}}\mathbf{v}_{b,\mathbf{p}}\cdot\mathbf{v}_{s,\mathbf{p}}, \\
    \mathcal{B}_s = \int_{\mathbf{p}} (-\partial_\varepsilon f^0_s)(\varepsilon_{s,\mathbf{p}}-\mu)\mathbf{v}_{s,\mathbf{p}}^2, \\
    \mathcal{C}_s = \int_{\mathbf{p}} (-\partial_\varepsilon f^0_s)\mathbf{v}_{s,\mathbf{p}}^2.
\end{gather}
To solve the constraint, we choose the solution
\begin{gather}
    \frac{1}{\tau_s} = \frac{s\left[\mathcal{A}_s +2\mathcal{B}_s + 2\mu\mathcal{C}_s  \right]}{\sum_{s=\pm}s\left[\mathcal{A}_s +2\mathcal{B}_s + 2\mu\mathcal{C}_s \right]} \frac{1}{\tau_0}
\end{gather}
where $\tau_0$ is some appropriate timescale, which we choose to be the Landau damping timescale $\tau_b$. In practice for our model, this quantity is estimated from a phase space averaging of the Landau damping, i.e.,
\begin{gather}\label{effective_Landau_damping}
    \tau^{-1}_b \equiv \frac{\int_{\mathbf{q}}\mathbf{v}_\mathbf{q}^2b_0[1+b_0] \gamma_\mathbf{q} }{\int_{\mathbf{q}}\mathbf{v}_\mathbf{q}^2b_0[1+b_0]}.
\end{gather}
Note that this integral implicitly includes the plasmon cutoff.
\subsection{Viscous transport}
To have particle-hole symmetry, i.e., $\eta(\mu) = \eta(-\mu)$ we need to use a slightly different collision matrix to that of Eq. (\ref{collision_matrix_thermoelectric}).
This is because the matrix 
\begin{gather}
    \mathcal{T}^{\eta}_{\gamma' \gamma} \equiv \frac{N_{\gamma'}}{30}\int_\mathbf{p}\left(\partial_\epsilon f_{\gamma}^0\right) Q^{\eta}_{\gamma,\mathbf{p}} Q^{\eta}_{\gamma',\mathbf{p}}\mathbf{v}_{\gamma, \mathbf{p}}\cdot\mathbf{v}_{\gamma', \mathbf{p}}
\end{gather}
satisfies $\mathcal{T}^{\eta}_{b\gamma}(\mu) = - \mathcal{T}^{\eta}_{b-\gamma}(-\mu)$ instead of $\mathcal{T}^{qq'}_{b\gamma}(\mu) =  \mathcal{T}^{qq'}_{b-\gamma}(-\mu)$, as was the case in the thermoelectric transform scenario. With this in mind, we can make a small alteration of Eq. (\ref{collision_matrix_thermoelectric}) so that the collision matrix used is 
\begin{gather} \label{collision_integral_visco}
C =
    \begin{pmatrix}
        \frac{1}{\tau_{+}} + \frac{1}{\tau_{\textrm{imp}}} && - \frac{1}{\tau_{-}}  && -\frac{1}{\tau_{b}} \\
        - \frac{1}{\tau_{+}}  &&  \frac{1}{\tau_{-}} + \frac{1}{\tau_{\textrm{imp}}}  && \frac{1}{\tau_{b}} \\ 
       0 && 0  && \frac{2}{\tau_{b}} 
    \end{pmatrix},
\end{gather}
where the element $C_{-b}$ had a sign flip. This collision matrix allows for a viscosity consistent with particle-hole symmetry.

\section{Helicons} \label{appendix_helicons}
The modes of the electric field can be found from solving the equation of motion for the electric field $\mathbf{e}$,
\begin{gather}\label{helicon_polarization_equation}
    [c^2q^2\delta_{ij} -c^2q_iq_j - \omega_h^2\varepsilon_{ij}(\omega)]e_{j} = 0,
\end{gather}
where $\varepsilon_{ij}$ is the dielectric tensor. In the limit $\omega \ll \omega_c$, it has the form
\begin{gather}
    \varepsilon_{ij} =  \begin{pmatrix}
        \varepsilon_\perp & ig & 0 \\
        -ig & \varepsilon_\perp & 0 \\
        0 & 0 & \varepsilon_\parallel
    \end{pmatrix}_{ij}
    + \frac{2c}{\pi}\epsilon_{ijk}b_k,
\end{gather}
where $b_k$ are the components of $\mathbf{b}$, the internodal separation in momentum space (we assume that the nodes are located at the same energy) between the pair of Weyl nodes, and
\begin{align}
    \varepsilon_\parallel = 1 - \frac{\omega_p^2}{\omega_c^2}, \quad
    \varepsilon_\perp = 1 + \frac{\omega_p^2}{\omega_c^2}, \quad
    g &= - \frac{\omega_p^2}{\omega \omega_c} .
\end{align}
The dielectric tensor can can be related to the polarization tensor components by
\begin{align}
    \Pi^{00} &= c^2q^i(\delta^{ij} - \varepsilon^{ij})q^j, \\
    \Pi^{ij} &= - \omega^2 c (\delta^{ij} - \varepsilon^{ij}), \\
    \Pi^{i0} &= \omega c(\delta^{ij} - \varepsilon^{ij})q^j, \\
    \Pi^{0j} &= -\omega c q^i(\delta^{ij} - \varepsilon^{ij}).
\end{align}
The dispersion of the modes that solve Eq. (\ref{helicon_polarization_equation}), i.e., the helicon modes, is $\omega_{h,\pm} = \pm \omega_h = \pm \frac{\omega_c}{\omega_p^2 + 2\alpha c b_3/\pi}q|q_3| + O(q^3)$, where we have distinguished the positive from the negative helicon frequencies. To the lowest order in $\mathbf{q}$, we can also find the electric field polarizations of eigenmodes $\mathbf{e}_\pm$ from Eq.(\ref{helicon_polarization_equation}),
\begin{equation}
    \mathbf{e}_{\pm} = \frac{1}{\sqrt{q_1^2 + q_3^2}}\begin{pmatrix}
        q_1^2 + q_3^2 \\
        q_1q_2 \mp i q|q_3| \\
      0
    \end{pmatrix} .
\end{equation}
These vectors are rotation invariant around the magnetic field axis, modulo a global phase. Since the formalism of the main text is written in terms of the 4-vector potential fields and not the electric field, we need to translate this polarization into the 4-vector polarization $u^\mu_\pm$, defined by
\begin{equation}
    A^\mu_Q = \sum_{s = \pm}\phi_{s,Q}u^\mu_{s}(\mathbf{q})\Theta(s\omega),
\end{equation}
where $\phi_{s,Q}$ is a scalar field which will go on to represent the helicon excitations. In the Coulomb gauge $(\mathbf{q} \cdot \mathbf{A} = 0)$, these electric field polarizations correspond to a polarization $A^\mu \propto (u^0,\mathbf{u})$ in the 4-vector potential of the form
\begin{align}
    u^0_\pm &= - \frac{i\mathbf{q}\cdot \mathbf{e}_{\pm}}{q^2}, \\
    \mathbf{u}_\pm &= \frac{i}{\omega_{h,\pm}}\left(\mathbf{e}_{\pm} -iu^0_\pm\mathbf{q}\right).
\end{align}
One can show that $u^\mu_{\pm}(\mathbf{q}) = [u^\mu_{\mp}(-\mathbf{q})]^*$. Hence, the kinetic term in the 4-vector potential Lagrangian (\ref{matter_EM_Lagrangian}) becomes 
\begin{gather}
    A^\mu_{-Q}(K_{\mu\nu} - \Re\Pi_{\mu\nu})A^\nu_Q 
   = Z^{-1}(\omega_h^2 - \omega^2)\phi_{-Q}\phi_Q,
\end{gather}
where \begin{gather}\label{aux_helicon_field}
    \phi_Q = e^{i\theta_\mathbf{q}}\phi_{+,Q}\Theta(\omega) + e^{-i\theta_\mathbf{q}}\phi_{-,Q}\Theta(-\omega)
\end{gather} and
\begin{equation}\label{helicon_renormalization}
    Z^{-1}_\mathbf{q} = \frac{(q_1^2 + q_2^2)q_3^2 c^2\omega_p^2}{q^2\omega_h^4}.
\end{equation}
The phase $\theta_\mathbf{q}$ will be defined shortly.

To look at interactions, we will consider only the scalar potential interaction with the fermion density. It can be restated as two interactions for positive and negative energy modes,
\begin{gather} \nonumber
e\int_P A^0_Q\Psi^\dagger_{P-Q}\Psi_P
= 
ec\sqrt{q_1^2 + q_2^2} \cdot \phi_Q \int_P\Psi^\dagger_{P-Q}\Psi_P.
\end{gather}
This form comes from the fact that
\begin{gather}
   u^0_\pm = \frac{-i\mathbf{q}\cdot \mathbf{e}_{\pm,\mathbf{q}}}{q^2} = \sqrt{q_1^2 + q_2^2} e^{\pm i\theta_\mathbf{q}}.
\end{gather}
Here, $\theta_{\mathbf{q}}$ is the phase used above in the definition of $\phi_Q$ in Eq. (\ref{aux_helicon_field}). It satisfies $\theta_\mathbf{q} = \theta_{-\mathbf{q}}$. Defining now the helicon field as $h_{Q} \equiv Z^{-\frac{1}{2}}_\mathbf{q} \phi_{Q}$, the kinetic term becomes the canonical one, whereas the interaction for the helicon field (the negative energy version is analogous) has the form
\begin{gather}
ec G_\mathbf{q}h_{Q}\int_P\Psi^\dagger_{P-Q}\Psi_P,
\end{gather}
where $G_\mathbf{q} \equiv Z^{\frac{1}{2}}_\mathbf{q} \sqrt{q_1^2+q_2^2}$. This is the expression given in Eq. (\ref{helicon_lagrangian}). Scattering processes will be proportional to the square of the leading factor,
\begin{equation}\label{helicon_scattering_guess}
    e^2\frac{Z_\mathbf{q}}{ 2\omega_h}\left\lvert -i\frac{\mathbf{q}\cdot \mathbf{e}_{\pm,\mathbf{q}}}{q^2}\right\rvert^2 = e^2\frac{\omega_h^4}{\omega_p^2c^2q_3^2}.
\end{equation}
Since $\omega_h$  converges to the constant value $\omega_c |q_3|/q$ (or alternatively $cq < \omega_p$) at large momenta, we see that when $\omega_h\rightarrow \omega_c |q_3|/q$ ($cq \rightarrow \omega_p$), the scattering will be suppressed by a factor of $\omega_c^4/\omega_p^4$. Therefore, we do not expect helicon-fermion scattering to be relevant in transport.

\bibliography{aapmsamp}
\end{document}